\documentclass[twocolumn,aps,superscriptaddress,showpacs,nofootinbib,floatfix]{revtex4}
\usepackage{epsfig,bm,feynmf}
\usepackage{graphicx}
\usepackage{amsmath}
\usepackage{dcolumn}

\usepackage{hyperref}
\hypersetup{
  colorlinks,
  unicode=True,
  linkcolor=Blue,
  citecolor=Blue,
  urlcolor=Blue,
}

\usepackage{graphicx}
\usepackage[usenames,dvipsnames,svgnames,table]{xcolor}

\usepackage{tikz-feynman,contour}
\tikzfeynmanset{compat=1.0.0}

\usepackage[normalem]{ulem}  

\renewcommand{\sout}{\bgroup \color{red} \ULdepth=-.5ex \ULset}

\begin{document}


\title{Thermal production of charm quarks in relativistic heavy-ion collisions}


\author{Taesoo Song}\email{t.song@gsi.de}
\affiliation{GSI Helmholtzzentrum f\"{u}r Schwerionenforschung GmbH, Planckstrasse 1, 64291 Darmstadt, Germany}

\author{Ilia Grishmanovskii}
\affiliation{Institut f\"ur Theoretische Physik, Johann Wolfgang Goethe-Universit\"at,Max-von-Laue-Str.\ 1, D-60438 Frankfurt am Main, Germany}

\author{Olga Soloveva}
\affiliation{Helmholtz Research Academy Hessen for FAIR (HFHF),GSI Helmholtz Center for Heavy Ion Physics. Campus Frankfurt, 60438 Frankfurt, Germany}
\affiliation{Institut f\"ur Theoretische Physik, Johann Wolfgang Goethe-Universit\"at,Max-von-Laue-Str.\ 1, D-60438 Frankfurt am Main, Germany}

\author{Elena Bratkovskaya}
\affiliation{GSI Helmholtzzentrum f\"{u}r Schwerionenforschung GmbH, Planckstrasse 1, 64291 Darmstadt, Germany}
\affiliation{Helmholtz Research Academy Hessen for FAIR (HFHF),GSI Helmholtz Center for Heavy Ion Physics. Campus Frankfurt, 60438 Frankfurt, Germany}
\affiliation{Institute for Theoretical Physics, Johann Wolfgang Goethe Universit\"{a}t, Frankfurt am Main, Germany}


\begin{abstract}
We investigate the thermal production of charm quarks in the strongly interacting quark-gluon plasma (sQGP) created in heavy-ion collisions at relativistic energies. Our study is based on the off-shell parton-hadron-string dynamics (PHSD) transport approach describing the full time evolution of heavy-ion collisions on a microscopic basis with hadronic and partonic degrees of freedom. The sQGP is realized within the effective dynamical quasi-particle model (DQPM) which is adjusted to reproduce the lattice QCD results for the thermodynamic observables of the sQGP. Relying on the fact that the DQPM successfully describes the spatial diffusion coefficients $D_s$ from the lQCD, which control the interaction of charm quarks with thermal partons (expressed in terms of strongly interacting off-shell quasiparticles), we evaluate the production of charm quark pairs through the rotation of Feynman diagrams such that the incoming charm quark and outgoing light parton in elastic scattering diagrams are exchanged. The charm quark annihilation is realized by detailed balance. We find that the number of produced thermal charm quark pairs strongly depends on the charm quark mass in the QGP. While for the heavy charm quarks of mass $m_c=1.8$ GeV it is subdominant compared to the primary charm production by binary nucleon-nucleon collisions at RHIC and LHC energies, the numbers of primary and thermal charm quarks become comparable for a smaller (bare) $m_c=1.2$ GeV. Compared with the experimental data on the $R_{\rm AA}$ of $D$-mesons in heavy-ion collisions at RHIC and LHC energies, it is more favorable for charm quarks in the QGP to gain additional mass due to thermal effects rather than to have a low bare mass.
\end{abstract}


\maketitle

\section{Introduction}

Heavy flavors (charm and beauty quarks/antiquarks) are one of the promising tomographic probes of the properties of hot and dense matter created in relativistic heavy-ion collisions. There have been extended experimental~\cite{STAR:2014wif,STAR:2018zdy,CMS:2017qjw,ALICE:2021rxa} and theoretical efforts (cf. Refs. ~\cite{Gossiaux:2009mk,He:2011qa,Uphoff:2012gb,Cao:2013ita,Song:2015sfa,Plumari:2017ntm,Cao:2019iqs,Beraudo:2022dpz} as well as the reviews \cite{Cao:2018ews,Rapp:2018qla,Xu:2018gux,Zhao:2023nrz}) in the last decade to improve our understanding of the production mechanisms of heavy flavours and their interactions in the quark-gluon plasma (QGP) and in the hadronic corona after hadronization. Since heavy flavors are massive, they are dominantly produced by primary nucleon-nucleon scatterings in heavy-ion collisions. Therefore, the number of produced heavy flavors scales with the number of nucleon binary collisions. However, heavy flavors can additionally be produced by partonic reactions in the QGP -- e.g., quark-antiquark annihilation or gluon-gluon fusion, which we denote by {\it "thermal"} production mechanism in the QGP even if it may happen at early reaction time, when the QGP does not equilibrate yet. With increasing collision energy, the temperature of produced QGP grows, and the possibility for thermal production of heavy flavor increases. This goes along in competition with the backward reactions of charm-anticharm quark annihilation. If the system achieves chemical equilibrium, the forward and backward reactions compensate each other, and the fugacity (which controls the deviation from a statistical phase-space distribution) of the charm quarks will approach 1 \cite{Mykhaylova:2022toe}. However, in realistic heavy-ion collisions, the system is far from equilibrium in the initial phase and also deviates from equilibrium during the fast expansion phase, not only chemically -- as shown in the Statistical Hadronization Model (SHM)~\cite{Andronic:2021erx} -- but also kinetically. Thus, one expects a violation of the balance between production/annihilation of charm-anticharm quarks produced in the QGP during the heavy-ion collisions.

Thermal production has been studied for more than three decades. Because charm production requires energetic scattering of partons, it was found that the early stage of heavy-ion collisions is important~\cite{Muller:1992xn,Geiger:1993py,Levai:1994dx}, and massive partons are more effective for it~\cite{Levai:1997bi}. In pQCD, calculations have been carried out up to next-to-leading order~\cite{Zhang:2007yoa}. These calculations commonly predicted that the thermal charm production would have sizable effects in heavy-ion collisions at high energies, such as those at the LHC. However, charm production, even at the LHC, is well explained in the SHM together with shadowing effects, which suppress charm production at midrapidity and at small $p_T$, corresponding to small $x$ in the parton distribution function. To understand this success, it is necessary to use reliable cross sections for charm production in parton scattering.

The interactions of heavy flavor with thermal partons in the QGP are partly known from the spatial diffusion coefficients of heavy flavor, which are provided by lQCD \cite{Banerjee:2011ra,Altenkort:2023eav}. The properties of a thermalized, strongly interacting QGP have been extensively studied with the Dynamical-QuasiParticle Model (DQPM) \cite{Peshier:2005pp,Cassing:2007nb,Cassing:2007yg,Berrehrah:2016vzw,Moreau:2019vhw,Soloveva:2020hpr}. The DQPM is an effective model that provides a microscopic picture of the thermalized QGP, reproducing its macroscopic properties, such as the equation of state (EoS) from lQCD, in terms of quasiparticles whose pole masses and spectral widths depend on temperature and baryon chemical potential~\cite{Moreau:2019vhw}. The DQPM is applied to parton interactions by using the leading-order Feynman diagrams and reproduces the spatial diffusion coefficient of heavy quark from lQCD in a wide range of temperatures \cite{Berrehrah:2014kba,Song:2019cqz}.

We stress that since the DQPM is based on propagator 2-particle irreducible representation, it is free from the pQCD divergences and necessity to introduce Debye masses for the regularisation of scattering amplitudes.
Indeed, the properties of quasiparticles expressed in terms of  complex self energies  -- with the real part related to the thermal paron masses and imaginary part to their widths -- are matched to reproduce the thermodynamic properties of lQCD at $\mu_B=0$ and extrapolates to finite $\mu_B$ using a scaling hypothesis. Thus, the DQPM propagators effectively includes the resummation effects since the self-energies are fitted to the lQCD data which a priori contain the full QCD solutions.

Based on this success, we extend the calculations to charm quark production in the QGP and in heavy-ion collisions by changing the incoming charm quark to the outgoing anticharm quark in the Feynman diagrams for the elastic scattering of a charm quark.

Since the charm quark mass is higher than the typical temperatures of the matter produced in heavy-ion collisions, the thermal production of charm quarks is usually ignored. However, the initial temperature of the sQGP, especially at the top energy of the LHC, is considerably high. Furthermore, the thermal masses of quarks and gluons increase with temperature in the DQPM, which is supported by the thermal quantum field theory~ \cite{Kapusta:2006pm,Bellac:2011kqa}. Both effects enhance the possibility of charm quarks being thermally produced in heavy-ion collisions.

The goal of our study is to estimate the amount of thermal charm-anticharm quark production by partonic interactions in the sQGP, in addition to the primary charm/anticharm production by initial binary nucleon-nucleon reactions in central heavy-ion collisions at RHIC and the LHC, and investigate their properties in comparison with the experimental data. Furthermore, we explore the balance between charm-anticharm quark production and annihilation in the expanding QGP matter created during heavy-ion collisions.

Assuming local thermal equilibrium, one can rely on hydrodynamics for the time evolution of nuclear matter in heavy-ion collisions. However, the initial thermalization time, from which hydrodynamics is applicable, is uncertain. On the other hand, this initial time is most important for the thermal/QGP production of charm due to its high temperature and/or high energy density. Moreover, in the initial phase of heavy-ion collisions, the system is out of equilibrium, so the application of hydrodynamical models at very early times is not possible.

In order to overcome these difficulties, we base our study of charm dynamics on a nonequilibrium transport approach, the parton-hadron-string dynamics (PHSD), which is a microscopic transport approach based on the first order gradient expansion of the Kadanoff-Baym equations, taking into account the off-shellness of particles for their strong interactions~\cite{Cassing:2008sv,Cassing:2009vt,Cassing:2008nn,Bratkovskaya:2011wp,Linnyk:2015rco}. Based on the DQPM for the description of the sQGP phase, it describes  the experimental data on open and hidden heavy flavors in relativistic heavy-ion collisions~\cite{Song:2015sfa,Song:2015ykw,Song:2016rzw}.

This paper is organized as follows: in Sec. \ref{sec:DQPM}, we briefly recall the DQPM. In Sec. \ref{sec:thermal_production}, the heavy quark production in the DQPM and the thermal production rate of charm quarks in the QGP are introduced. In Sec. \ref{sec:HIC}, we study the thermal charm production in heavy-ion collisions including charm quark pair annihilation by the detailed balance. The dependence on the charm quark thermal mass is also investigated in comparison with experimental data. Furthermore, we study the dependence of the charm diffusion coefficient $D_s$ on the choice of the charm quark mass $m_c$. Finally, a summary is given in Sec. \ref{sec:summary}.

\section{Dynamical Quasiparticle Model}
\label{sec:DQPM}

Here, we briefly recall the basic ideas of the Dynamical Quasiparticle Model (DQPM) \cite{Peshier:2005pp, Cassing:2007nb, Cassing:2007yg, Berrehrah:2016vzw, Moreau:2019vhw, Soloveva:2020hpr}, which is an effective model for the  description of the sQGP in terms of strongly interacting quasiparticles (quarks and gluons) matched to reproduce  the results of lattice QCD calculations in thermal equilibrium and at vanishing chemical potential. The quasiparticles are characterized by single-particle (two-point) Green's functions, i.e., "dressed" propagators:
\begin{equation}
    G^{R}_j (\omega, \vec{p}) = \frac{1}{\omega^2 - \vec{p}^2 - M_j^2 + 2 i \gamma_j \omega}
    \label{eq:propdqpm}
\end{equation}
for quarks, antiquarks, and gluons ($j = q,\bar{q},g$), using $\omega=p_0$ for energy, the widths $\gamma_{j}$, the masses $M_{j}$, and the complex self-energies for gluons $\Pi = M_g^2-2i \omega \gamma_g$ and for (anti)quarks $\Sigma_{q} = M_{q}^2 - 2 i \omega \gamma_{q}$, where the real part of the self-energies is associated with dynamically generated thermal masses, while the imaginary part provides information about the lifetime and reaction rates of the particles.

We use an ansatz for the pole masses $M_{j}(T,\mu_q)$ and widths $\gamma_{j}(T,\mu_q)$ as functions of the temperature $T$ and the quark chemical potential $\mu_q$, based on the HTL thermal mass in the asymptotic high-temperature regime  \cite{Bellac:2011kqa,Linnyk:2015rco}:
\begin{equation}
    M^2_{g}(T,\mu_q)=\frac{g^2(T,\mu_q)}{6}\left(\left(N_{c}+\frac{1}{2}N_{f}\right)T^2
    +\frac{N_c}{2}\sum_{q}\frac{\mu^{2}_{q}}{\pi^2}\right),
    \label{eq:Mg}
\end{equation}
for gluons, and for quarks (antiquarks) by
\begin{equation}
    M^2_{q(\bar q)}(T,\mu_q)=\frac{N^{2}_{c}-1}{8N_{c}}g^2(T,\mu_q)\left(T^2+\frac{\mu^{2}_{q}}{\pi^2}\right),
    \label{eq:Mq}
\end{equation}
where $N_{c}\ (=3)$ stands for the number of colors, and $N_{f}\ (=3)$ denotes the number of light flavors. 

The widths $\gamma_j$ of quasiparticles are taken in the form \cite{Linnyk:2015rco}:
\begin{equation}
    \gamma_{j}(T,\mu_\mathrm{B}) = \frac{1}{3} C_j \frac{g^2(T,\mu_\mathrm{B})T}{8\pi}\ln\left(\frac{2c_m}{g^2(T,\mu_\mathrm{B})}+1\right).
    \label{eq:widths}
\end{equation}
Here, $c_m = 14.4$ is related to a magnetic cutoff, which is an additional parameter in the DQPM, while $C_q = \dfrac{N_c^2 - 1}{2 N_c} = 4/3$ and $C_g = N_c = 3$ are the QCD color factors for quarks and gluons, respectively. We also assume that all (anti)quarks have the same $T$ dependence for the width. 

The thermal properties of quasiparticles and their interactions (defined via transport coefficients) strongly depend on the coupling constant $g$, which accounts for nonperturbative effects and enters the definition of the masses and widths of quasiparticles -- Eqs. \eqref{eq:Mq}, \eqref{eq:widths}. In the DQPM, $g^2$ is extracted from lQCD data on the entropy density $s$ by a parametrization method introduced in Ref. \cite{Berrehrah:2015vhe}, using the scaling of the ratio $s(T,g^2)/T^3$ versus $T$ for a given value of $g^2$:
\begin{equation}
    g^2(T,\mu_B=0) = d \left( (s(T,0)/s^{QCD}_{SB})^e - 1 \right)^f.
    \label{eq:coupling_DQPM}
\end{equation}
Here $d = 169.934, e = -0.178434$, and $f = 1.14631$ are the dimensionless parameters obtained by adjusting the quasiparticle entropy density $s(T,\mu_B=0)$ to the lQCD data provided by the BMW Collaboration \cite{Borsanyi:2012cr,Borsanyi:2013bia}, and $s^{QCD}_{SB} = 19/9 \pi^2T^3$ is the Stefan-Boltzmann entropy density. The extension of the coupling constant to finite baryon chemical potential $\mu_B$ is realized using a scaling hypothesis \cite{Cassing:2008nn} that works up to $\mu_B \approx 500$ MeV. 

\begin{figure}[t!]
    \includegraphics[width=8. cm]{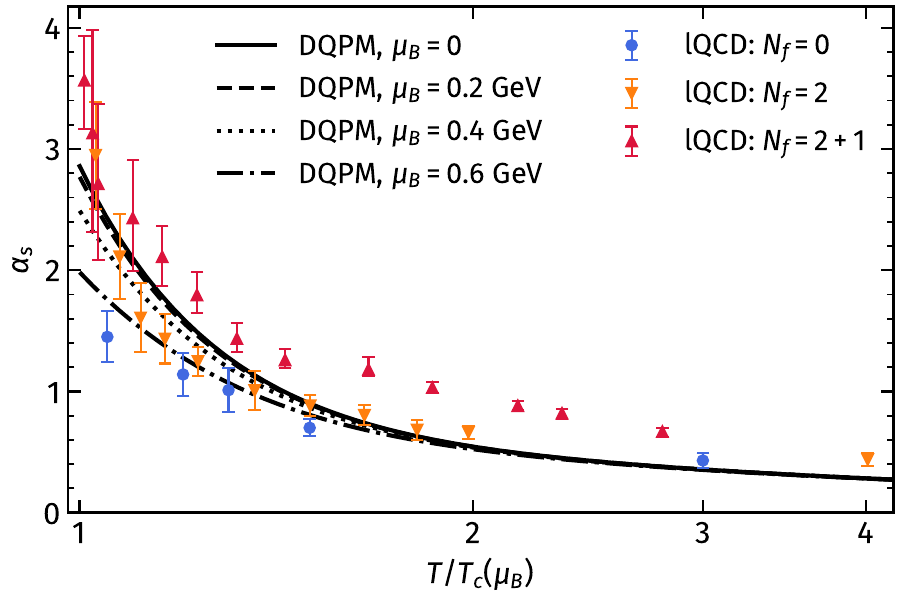}
    \caption{
        Strong coupling $\alpha_s$ as a function of scaled temperature at $\mu_B=$ 0, 0.2, and 0.4 GeV, compared with lattice results.
    }
    \label{alphas}
\end{figure}

As shown in Fig.~\ref{alphas}, the resulting $\alpha_s$ starts with a small value at high temperature but increases with decreasing temperature and reaches 2.9 for $\mu_B=0$ and 2.0 for $\mu_B=0.6$ GeV at $T_c$~\cite{Moreau:2019vhw,Grishmanovskii:2023gog}, which is consistent with the lattice QCD results~\cite{Kaczmarek:2004gv, Kaczmarek:2005ui, Kaczmarek:2007pb}.

By comparison of the entropy density -- computed within the DQPM framework -- to the lQCD data, one can fix the few parameters used in the ansatz for the quasiparticle masses and widths. Hence, the DQPM provides the quasiparticle properties, including dressed propagators and coupling constants. The coupling $g$ and propagators $G_j^R$ allow for computing scattering amplitudes, cross sections, and transport coefficients of quarks and gluons in sQGP -- cf. Refs. \cite{Berrehrah:2013mua,Moreau:2019vhw,Grishmanovskii:2023gog}.

\section{Thermal production of charm quarks in the QGP}\label{DQPM}
\label{sec:thermal_production}

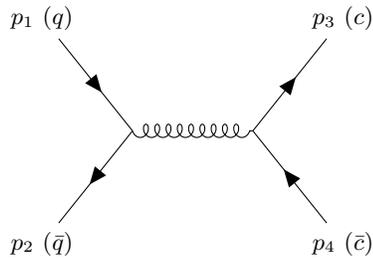
\begin{figure}[ht!]
    \centering
    \begin{tikzpicture}
    \begin{feynman}
    \vertex (b) at (-0.8,0.);
    \vertex (a) at (-2.,1.5) {$p_1$ ($q$)};
    \vertex (c) at (-2.,-1.5) {$p_2$ ($\bar{q}$)};
    \vertex (e) at (0.8,0.);
    \vertex (f) at (2.,1.5) {$p_3$ ($c$)};
    \vertex (d) at (2.,-1.5) {$p_4$ ($\bar{c}$)};
    \diagram* {
        (a) -- [fermion] (b) -- [fermion] (c),
        (d) -- [fermion] (e) -- [fermion] (f),
        (b) -- [gluon] (e),
       };
    \end{feynman}
    \end{tikzpicture}
    \caption{
        Charm quark pair production by annihilation of a quark and antiquark pair.
    }
    \label{qqbar-fig}
\end{figure}

The thermal production of charm quarks in a QGP is closely related to the elastic scattering of charm quarks in the QGP. Figures \ref{qqbar-fig} and \ref{gg-fig} show the Feynman diagrams for thermal production from quark-antiquark annihilation and two-gluon fusion, respectively. They are obtained by rotating the Feynman diagrams for charm quark elastic scattering with light (anti)quarks or gluons in Ref.~\cite{Berrehrah:2013mua}. The calculations of the Feynman diagrams are performed by exchanging the incoming charm quark with the outgoing light quark or a gluon in the elastic scattering diagrams. For example, the squared scattering amplitudes are related to each other as follows:
\begin{eqnarray}
&&|M|_{q\bar{q}\rightarrow c\bar{c}}^2(p_1,p_2;p_3,p_4)\nonumber\\
&&~~~~~=|M|_{qc\rightarrow qc}^2(p_1,-p_4;-p_2,p_3),\\
&&|M|_{gg\rightarrow c\bar{c}}^2(p_1,p_2;p_3,p_4)\nonumber\\
&&~~~~~=-|M|_{gc\rightarrow gc}^2(p_1,-p_4;-p_2,p_3),
\end{eqnarray}
where the second equation has an overall minus sign because only one (odd number) fermion is changed to an anti-fermion.

\begin{figure}[ht!]
\centerline{
  \begin{tikzpicture}
    \begin{feynman}
      \vertex (b) at (-0.8,0.);
      \vertex (a) at (-2.,1.5) {$p_1$ };
      \vertex (c) at (-2.,-1.5) {$p_2$ };
      \vertex (e) at (0.8,0.);
      \vertex (f) at (2.,1.5) {$p_3$ ($c$)};
      \vertex (d) at (2.,-1.5) {$p_4$ ($\bar{c}$)};
      \diagram* {
        (a) -- [gluon] (b) -- [gluon] (c),
        (d) -- [fermion] (e) -- [fermion] (f),
        (b) -- [gluon] (e),
       };
    \end{feynman}
  \end{tikzpicture}
}
\vspace{2em}
\centerline{
  \begin{tikzpicture}
    \begin{feynman}
      \vertex (b) at (0.,1.5);
      \vertex (a) at (-1.5,1.5) {$p_1$};
      \vertex (c) at (-1.5,-1.5) {$p_2$};
      \vertex (e) at (0.,-1.5);
      \vertex (f) at (1.5,1.5) {$p_3$ ($c$)};
      \vertex (d) at (1.5,-1.5) {$p_4$ ($\bar{c}$)};
      \diagram* {
        (d) -- [fermion] (e) -- [fermion] (b) -- [fermion] (f),
        (b) -- [gluon] (a),
        (e) -- [gluon] (c),        
       };
    \end{feynman}
  \end{tikzpicture}
\quad
  \begin{tikzpicture}
    \begin{feynman}
      \vertex (b) at (0.,1.5);
      \vertex (a) at (-1.5,1.5) {$p_1$};
      \vertex (c) at (-1.5,-1.5) {$p_2$};
      \vertex (e) at (0.,-1.5);
      \vertex (f) at (1.5,1.5) {$p_3$ ($c$)};
      \vertex (d) at (1.5,-1.5) {$p_4$ ($\bar{c}$)};
      \diagram* {
        (d) -- [fermion] (e) -- [fermion] (b) -- [fermion] (f),
        (b) -- [gluon] (c),
        (e) -- [gluon] (a),        
       };
    \end{feynman}
  \end{tikzpicture}
}
  \caption{Charm quark pair production by the fusion of two gluons.}
\label{gg-fig}
\end{figure}

We note that Figs.~\ref{qqbar-fig} and \ref{gg-fig} are not simple leading-order diagrams in the DQPM, as the exchanged quark and gluon have dressed masses and widths from the interactions in the QGP. The mass and width depend on temperature and baryon chemical potential, and are extracted from the lQCD equation of state \cite{Moreau:2019vhw}.

Since charm quark elastic scattering in the DQPM is quantitatively supported by lQCD in terms of the spatial diffusion coefficients~\cite{Song:2019cqz}, the thermal production of charm quarks in the DQPM is expected to be reliable.
In the case of elastic scattering the dressed mass and spectral width of light partons play important role to remove singularities and reduce forward scattering compared to the scattering of charm quark with massless partons~\cite{Berrehrah:2013mua}.
However, this role is not manifested in the case of charm production, because $t-$channel in the elastic scattering turns to $s-$channel in the charm production, which does not have singularity, and the forward scattering is suppressed in the charm production even for the massless light partons.

\begin{figure}[t!]
    \includegraphics[width=8.6 cm]{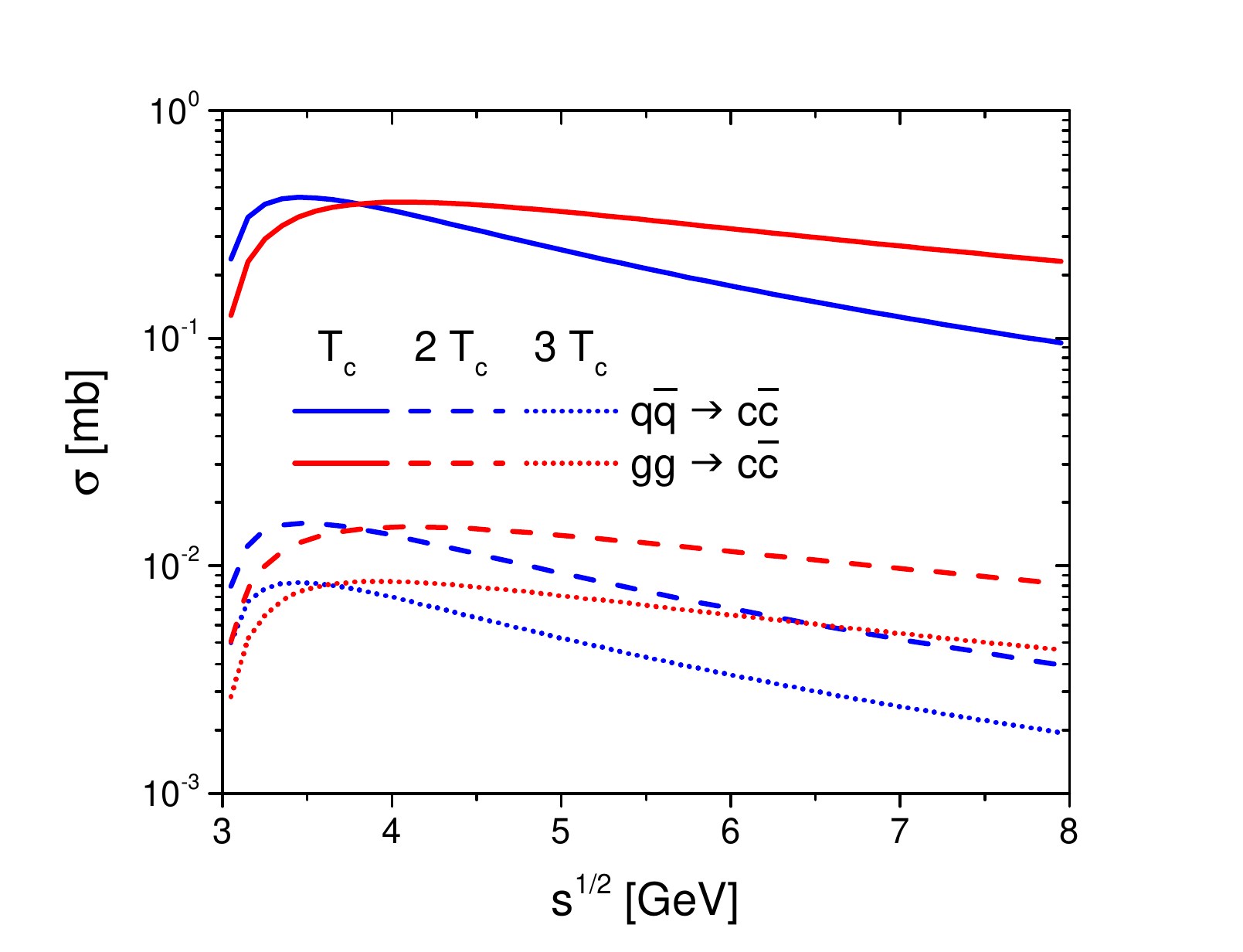}
    \includegraphics[width=8.6 cm]{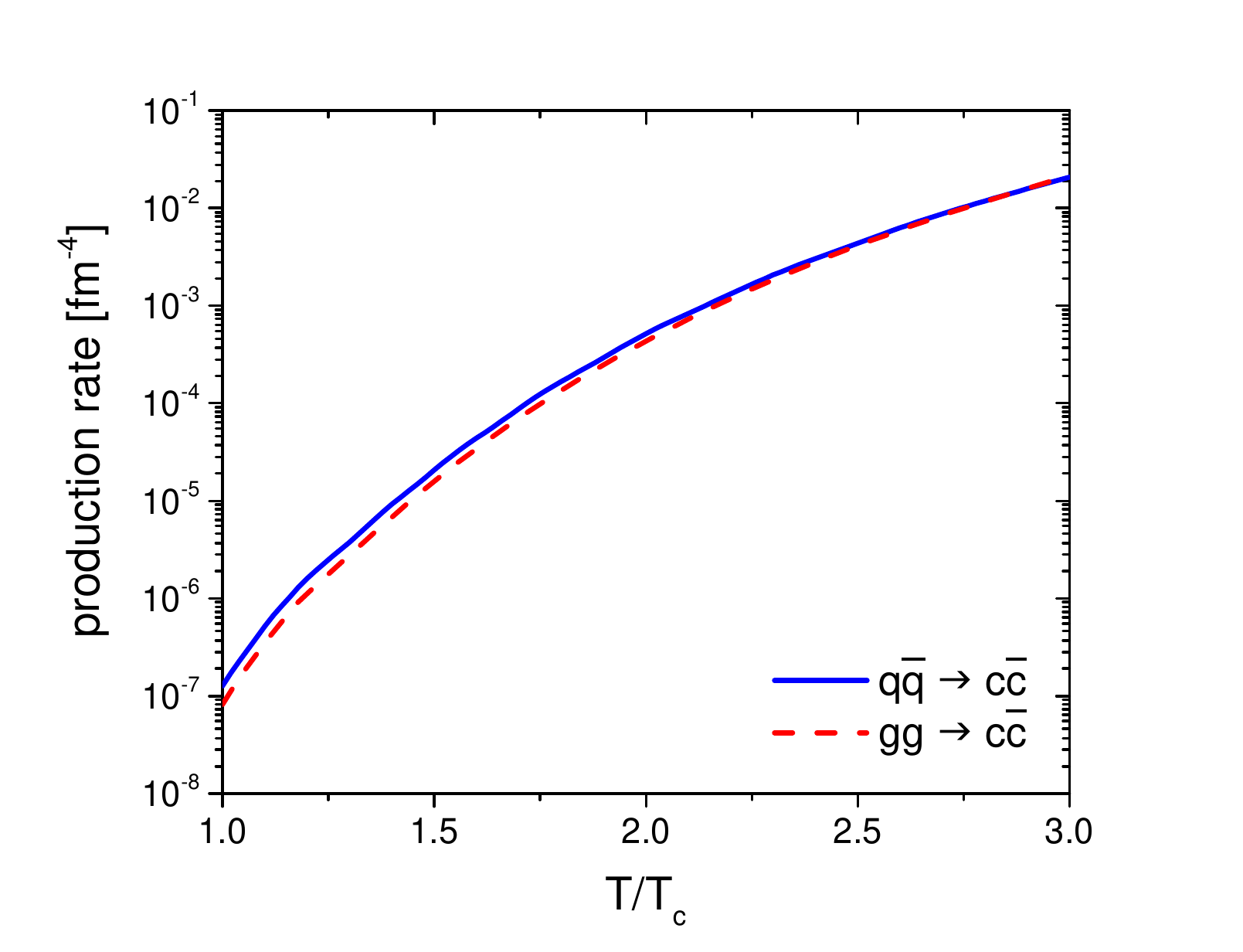}
    \caption{(Top) scattering cross sections of a light quark-antiquark pair and of two gluons to produce a charm quark pair as a function of scattering energy $\sqrt{s}$ at $T=T_c$, 2 $T_c$, and 3 $T_c$; (bottom) the production rate of charm quark pairs in the QGP as a function of temperature. The charm quark mass is taken to be 1.5 GeV.}
    \label{sigma-rate-fig}
\end{figure}

Figure \ref{sigma-rate-fig} shows the scattering cross sections for charm production from quark-antiquark annihilation and two-gluon fusion as a function of scattering energy at the temperatures of 1, 2, and 3 $T_c$. The charm quark mass is taken to be 1.5 GeV. Both cross sections increase rapidly above the threshold energy and then decrease. Since the strong coupling has a maximum at $T_c$ and decreases with increasing temperature, the production cross sections are largest at $T_c$ and much smaller at 2 and 3 $T_c$. The lower panel shows the production rate of charm quark pairs per unit time and unit volume, defined by
\begin{eqnarray}
\Gamma_{q\bar{q}\rightarrow c\bar{c}}= \sum_{q=u,d,s}\int dm_1 \rho_q(m_1)\int dm_2 \rho_{\bar{q}}(m_2)\nonumber\\
\times \int \frac{d^3k_1 d^3k_2}{(2\pi)^6 } f_q(k_1,m_1) f_{\bar{q}}(k_2,m_2)v_{q\bar{q}}\sigma_{q\bar{q}\rightarrow c\bar{c}},\\
\Gamma_{gg\rightarrow c\bar{c}}=\int dm_1 \rho_g(m_1)\int dm_2 \rho_g(m_2)~~~~~~~\nonumber\\
\times \int \frac{d^3k_1 d^3k_2}{(2\pi)^6 } f_g(k_1,m_1) f_g(k_2,m_2)v_{gg}\sigma_{gg\rightarrow c\bar{c}};
\label{rate}
\end{eqnarray}
where $\rho_i$ and $f_i$ are the spectral functions and momentum distribution function of parton $i$ at temperature $T$, respectively; $v_{ij}$ is the relative velocity of partons $i$ and $j$, and $\sigma$ is the scattering cross section.

Although the production cross section has a maximum at $T_c$, the production rate is quite small because only a few reactions of quark-antiquark annihilation or two-gluon fusion are above the threshold energy for charm pair production at low temperature. When comparing the two channels, quark-antiquark annihilation has a slightly larger production rate than two-gluon fusion, although the difference is small.
We note that the off-shell effect of light partons enhances the charm production rate from gluon fusion by about 7\%, while the enhancement of the rate from quark annihilation is less than 1\%.

We also note that our $2\to 2$ production rates are 
in line  with those from Ref.~\cite{Zhang:2007yoa} which considers not only 2$\rightarrow$2 reactions but also next-to-leading order 2$\rightarrow$3 reactions ($q+\bar q \to c + \bar c + g$ and  $g+g \to c + \bar c + g$) within the pQCD framework, which give a visible contribution to the production of charm pairs. 
However, for the case of non-perturbative QCD degrees-of-freedom of the DQPM, it has been shown in Ref. \cite{Grishmanovskii:2023gog} that the contribution  of the $2\to 3$ processes for the emission of a massive gluon to the thermal rate in the equilibrated QGP is suppressed contrary to the elastic $2\to 2$ scattering (cf.  Fig. 16 of Ref. \cite{Grishmanovskii:2023gog}). 
Thus, we expect that the processes  $q+\bar q \to c + \bar c + g$ and  $g+g \to c + \bar c + g$ would be even more strongly suppressed due to the large mass of emitted gluon in thermal equilibrium. 
To clarify this issue quantitatively would require a further study.
Moreover, since in Ref.~\cite{Zhang:2007yoa} the smaller charm quark mass is used and 2$\rightarrow$3 reactions are included, the obtained production rates are larger than in this study.
However, the rates are applied in Ref.~\cite{Zhang:2007yoa} only after the initial thermalization time of a fireball, which will be overcome in our study by using the production cross sections directly, as shown in the next section.

\section{Thermal production of charm quark pairs in heavy-ion collisions}
\label{sec:HIC}

Now, we compute the thermal production of charm pairs in heavy-ion collisions at RHIC and LHC energies. For this purpose, the PHSD is used to simulate heavy-ion collisions~\cite{Song:2015sfa,Song:2015ykw}, where charm is produced only through initial hard scattering of nucleons. The energy-momentum of (anti)charm quarks is provided by the PYTHIA event generator~\cite{Sjostrand:2006za}, in which the transverse momentum and rapidity are tuned to match those in the Fixed-Order Next-to-Leading Logarithm (FONLL) calculations~\cite{Cacciari:2012ny}. In heavy-ion collisions, the charm energy-momentum distribution is also modified due to the (anti)shadowing effects, which are realized in PHSD by EPS09~\cite{Eskola:2009uj}. We recall that charm production is suppressed at low transverse momentum and at midrapidity, which arises from small $x$ of the parton distribution function~\cite{Song:2015ykw}.

Since neither thermal production nor annihilation of charm quarks is considered, the number of charm quarks is conserved in heavy-ion collisions. However, by including the processes in Figs.~\ref{qqbar-fig} and \ref{gg-fig} for charm production as well as charm annihilation in the QGP, the charm number is not conserved anymore.

\begin{figure} [h]
    \centerline{
    \includegraphics[width=8.6 cm]{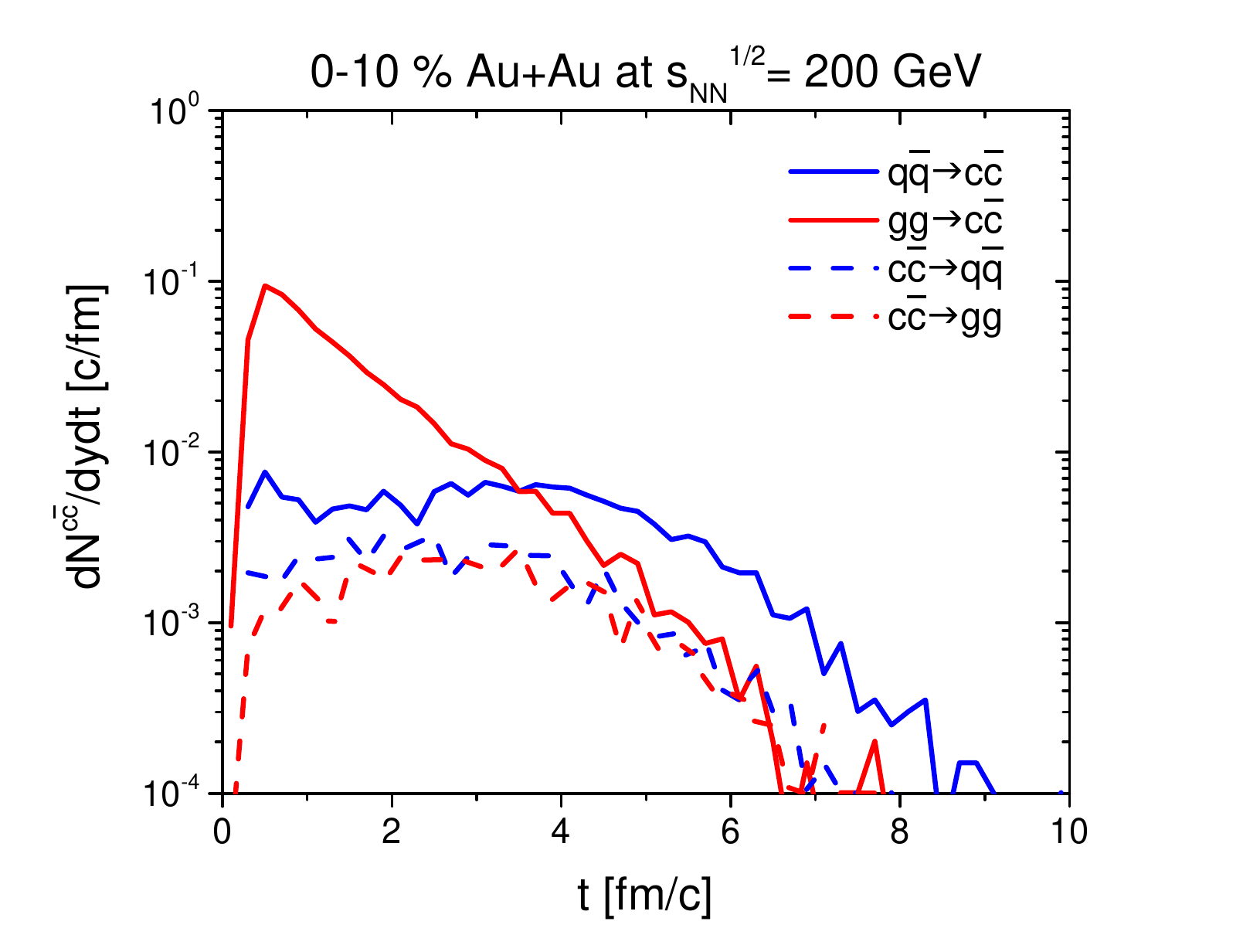}}
    \centerline{
    \includegraphics[width=8.6 cm]{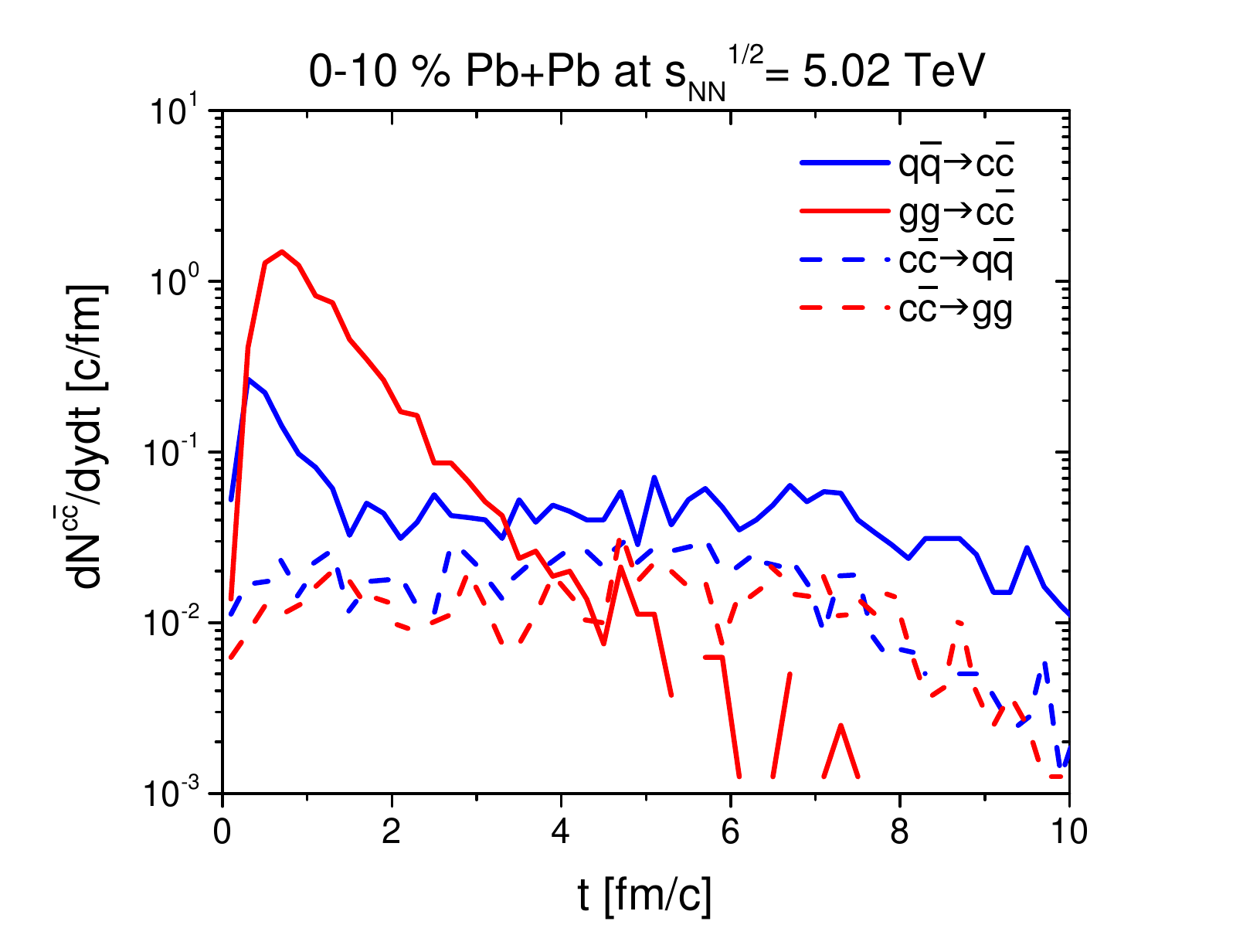}}    \caption{
        The PHSD results for the numbers of charm pair production and annihilation per rapidity bin at midrapidity as a function of time (upper) in central Au+Au collisions at $\sqrt{s_{\rm NN}}=$ 200 GeV and (lower) in central Pb+Pb collisions at $\sqrt{s_{\rm NN}}=$ 5.02 TeV. The charm quark mass is taken to be 1.5 GeV. 
    } 
    \label{reactions}
\end{figure}

\begin{figure} [ht!]
    \centerline{
    \includegraphics[width=8.6 cm]{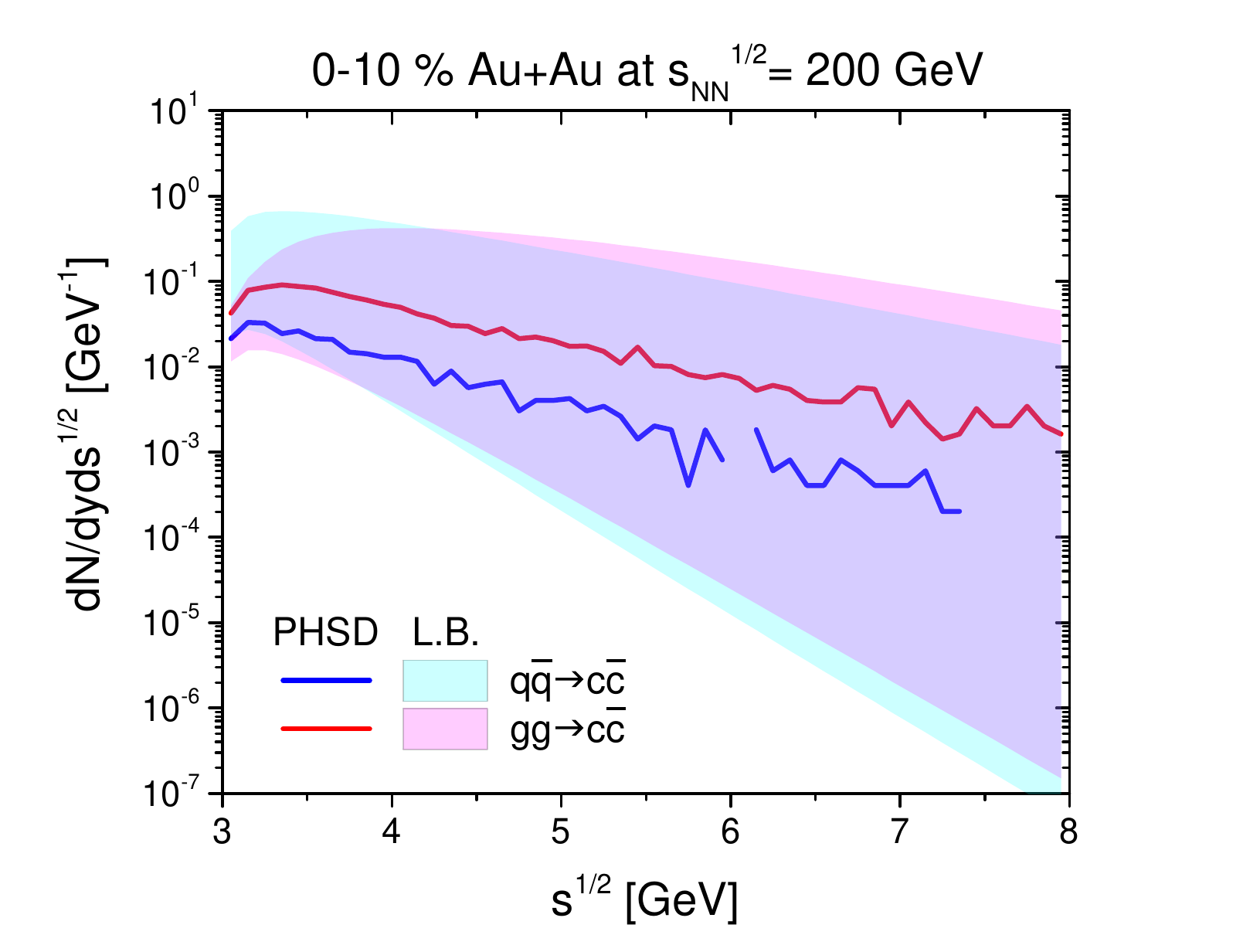}}
    \centerline{
    \includegraphics[width=8.6 cm]{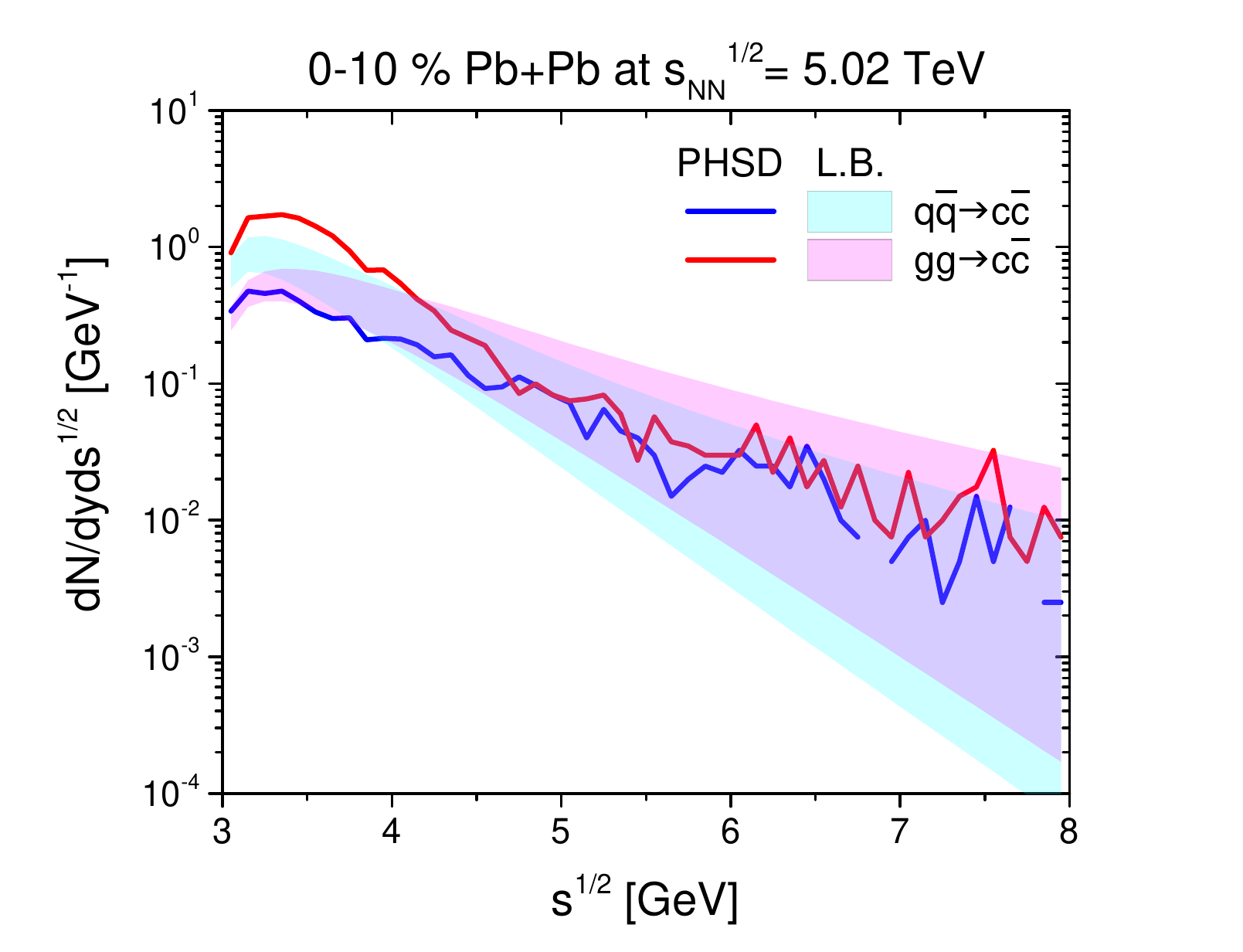}}
    \caption{
        The distributions of scattering energy $\sqrt{s}$ for thermal charm production per rapidity bin at midrapidity for PHSD and for the linearized Boltzmann approach in the PHSD assuming local thermal equilibrium (upper) in central Au+Au collisions at $\sqrt{s_{\rm NN}}=$ 200 GeV and (lower) in central Pb+Pb collisions at $\sqrt{s_{\rm NN}}=$ 5.02 TeV. The upper and lower limits of the bands indicate the thermal charm production from $t=$ 0 $fm/c$ and 0.5 $fm/c$, respectively, in the linearized Boltzmann approach. The charm quark mass is taken to be 1.5 GeV. 
    } 
    \label{dists}
\end{figure}

Figure \ref{reactions} shows the number of thermally produced and annihilated charm quark pairs per rapidity bin at midrapidity as a function of time in central Au+Au collisions at $\sqrt{s_{\rm NN}}=$ 200 GeV and in central Pb+Pb collisions at $\sqrt{s_{\rm NN}}=$ 5.02 TeV for a charm quark mass of 1.5 GeV. One can see that the thermal production of charm dominantly takes place in the early stage of heavy-ion collisions, where the temperature is high and partons have larger energy to produce a charm quark pair. If the partonic matter is completely thermalized, the contributions from quark-antiquark annihilation and from two-gluon fusion will be similar. However, the initial stage of heavy-ion collisions is far from equilibrium. In the literature, different types of off-equilibrium are quoted \cite{Song:2019cqz}, for example, the chemical off-equilibrium, the kinetic off-equilibrium, the anisotropic off-equilibrium, and the off-equilibrium in parton spectral functions. The PHSD accounts for all of them. 

In the PHSD, initial partons are produced by string melting when the local energy density exceeds a critical value of approximately 0.5 GeV/fm$^3$. The mass distributions of quarks and gluons follow the spectral functions from the DQPM through scatterings. We note that the initial mass distributions of quarks and gluons differ from the thermalized spectral distributions of the DQPM. In the PHSD, gluons are initially more massive, but (anti)quarks are less massive compared to the equilibrium spectral functions of the DQPM. Thus, the thermal production of charm quark pairs is dominated by two-gluon fusion in the early stage of heavy-ion collisions, as shown in Fig.~\ref{reactions}. We also note that initial partons, which are produced by string melting in the PHSD, have a formation time that is proportional to $E/m_T^2$, with $E$ and $m_T$ being energy and transverse mass of the parton, respectively; the parton does not interact before the formation time has passed. This is why the charm production is suppressed at a very early time ($\le 0.2$ fm/c) in Fig.~\ref{reactions}.

When the local temperature decreases and approaches $T_c$, all gluons split into quark and antiquark pairs, followed by quark coalescence to form mesons and (anti)baryons. Therefore, two-gluon fusion is suppressed, and quark-antiquark annihilation becomes the dominant channel for thermal charm quark production, although the production rate is much smaller than in the initial gluon-gluon fusion. The number of produced thermal charm pairs per unit rapidity at midrapidity is about 0.15 in central Au+Au collisions at 200 GeV and 2.1 in central Pb+Pb collisions at 5.02 TeV.

On the other hand, charm quark annihilation, which is realized by detailed balance, contributes similarly to the quark-antiquark production and two-gluon production as shown by the dotted lines in Fig.~\ref{reactions}, but these contributions are less than those of the charm production channels.

Figure \ref{dists} shows the scattering energy distribution, which produces thermal charm quark pairs at midrapidity in central heavy-ion collisions at RHIC and LHC energies. To observe the non-equilibrium effects of partonic matter on charm production, the results from PHSD are compared with those from the linearized Boltzmann approach \cite{Song:2020tfm}, shown as colored bands for quark-antiquark annihilation and two-gluon fusion channels. The upper limit of the band indicates thermal charm production from $t=$ 0 fm/$c$, assuming thermal equilibrium as soon as the two nuclei pass through each other, and the lower limit is for $t=$ 0.5 fm/$c$, which is a typical initial thermalization time for hydrodynamical simulations at RHIC and LHC energies. One can see that the results from the linearized Boltzmann approach are highly sensitive to the initial thermalization time because the temperature in the initial stage is high and more effective to produce thermal charm quarks. The colored bands are wider at RHIC because the lifetime of the QGP is relatively short, and the initial time period between $t=$ 0 and 0.5 fm/$c$ is more important than at the LHC. The broad bands in the figure show the difficulty of studying thermal charm production in a hydrodynamical approach since in these calculations charm production before the initial thermalization time is not included.

\begin{figure} [h!]
    \centerline{
    \includegraphics[width=8.6 cm]{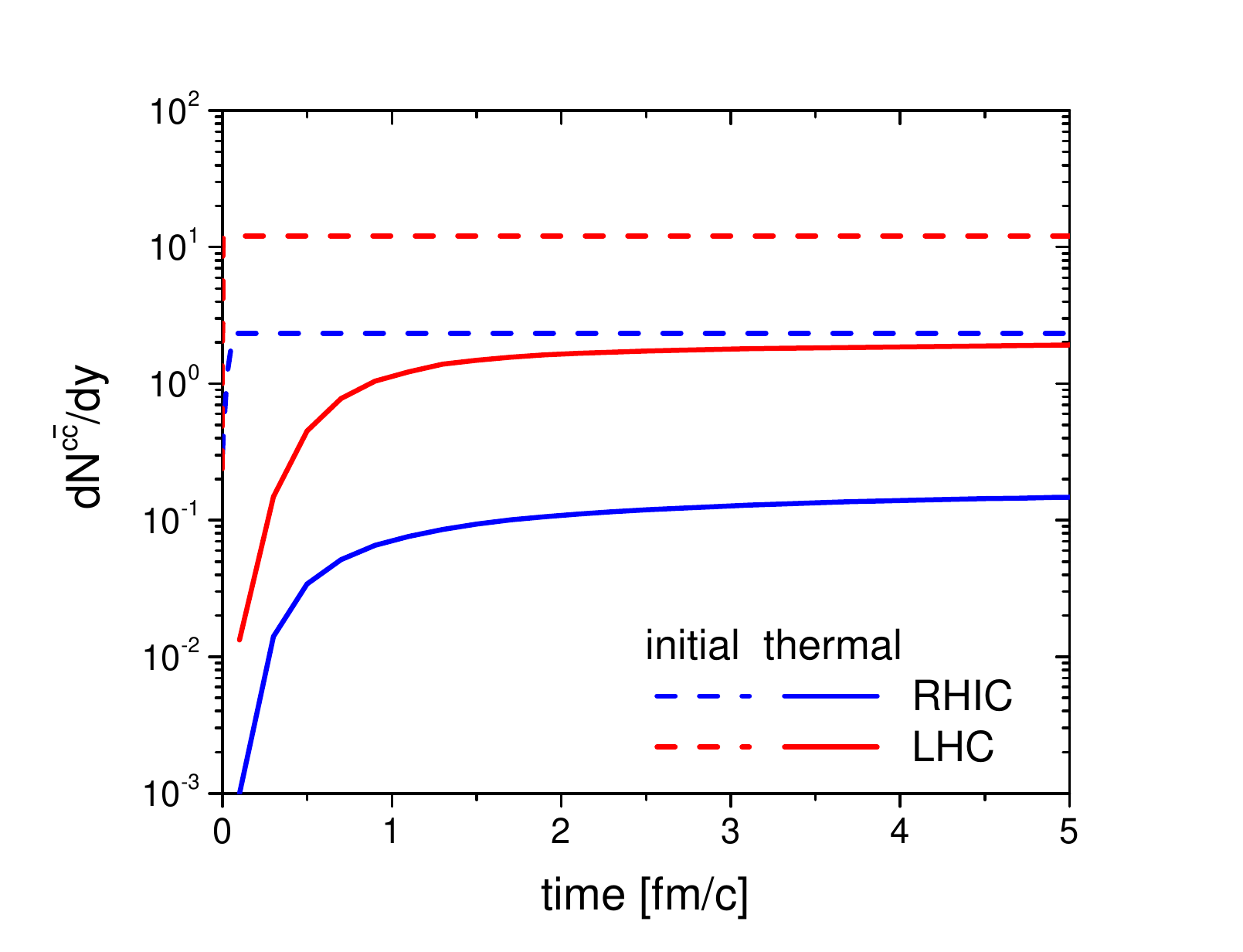}}
    \centerline{
    \includegraphics[width=8.6 cm]{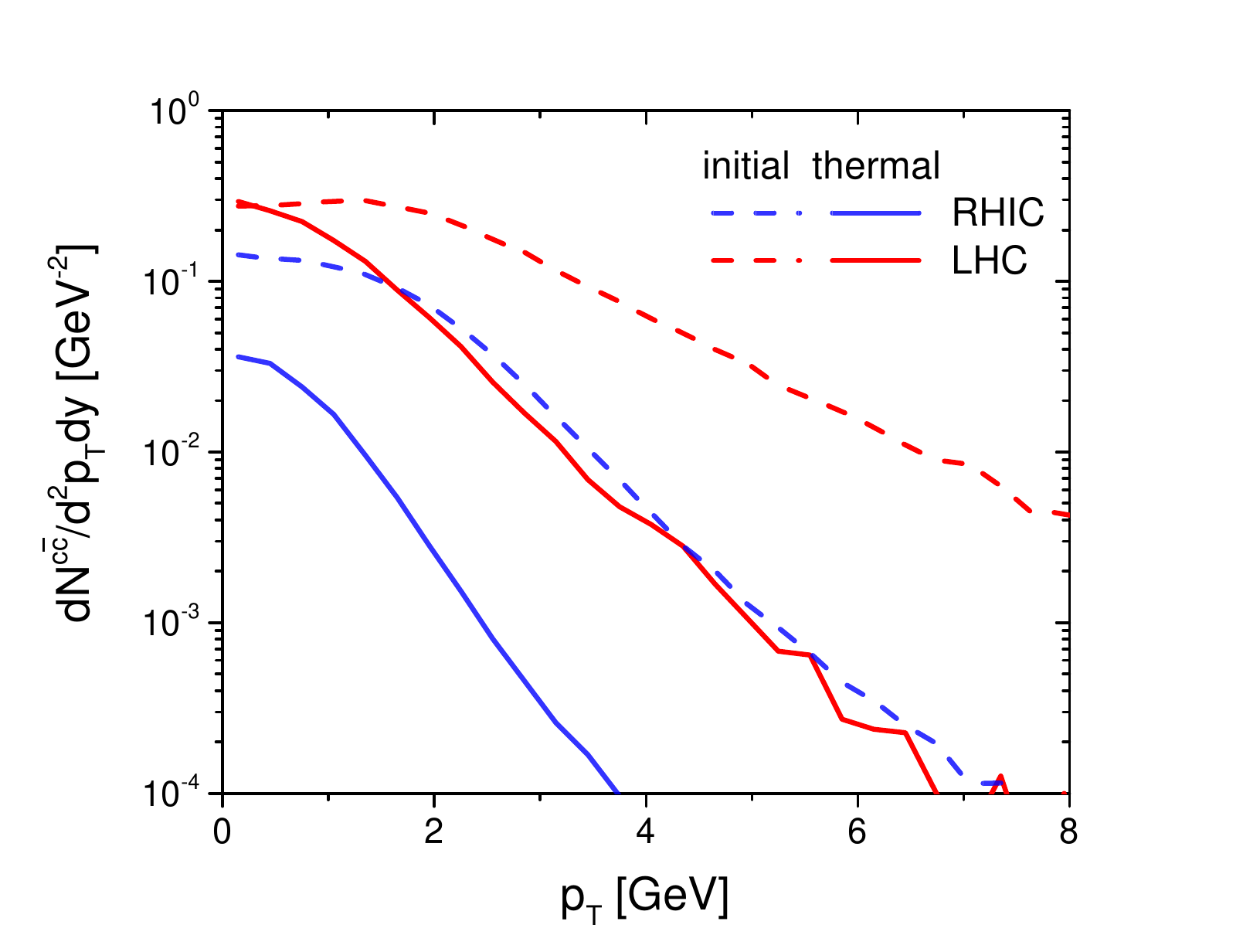}}
    \caption{
        (Upper) The PHSD results for the numbers of charm quark pairs produced initially and thermally at midrapidity as a function of time and (lower) their $p_T$ spectra in central Au+Au collisions at $\sqrt{s_{\rm NN}}=$ 200 GeV and in central Pb+Pb collisions at $\sqrt{s_{\rm NN}}=$ 5.02 TeV from the PHSD. The charm quark mass is taken to be 1.5 GeV. 
    }
    \label{dist-fig}
\end{figure}

In the upper panel of Fig.~\ref{dist-fig}, we show the number of initially and thermally produced charm quark pairs at midrapidity as a function of time in central Au+Au and Pb+Pb collisions at RHIC and the LHC. The initial production, caused by primary nucleon-nucleon binary scattering, takes place shortly after the two nuclei overlap. On the other hand, thermal production takes time, and its number increases gradually, although most are produced before $t\simeq 1-2$ fm/$c$. One can see that the number of thermal charm quarks at the LHC is comparable to that of initial charm quarks at RHIC for $m_c$= 1.5 GeV.

The lower panel shows the $p_T$ spectra of initially produced charm quark pairs and those of thermal charm quark pairs for $m_c=$ 1.5 GeV at midrapidity in central Au+Au collisions at $\sqrt{s_{\rm NN}}=$ 200 GeV and in central Pb+Pb collisions at $\sqrt{s_{\rm NN}}=$ 5.02 TeV. We note that these are spectra at the production time and that the interactions in the QGP are not included. Comparing the solid and dashed lines, the spectra of the thermal charm are softer than those of the initial hard charm pairs. Counting the numbers, the number of thermal charm quarks at midrapidity is about 6.5\% of the number of initially produced charm quarks at the same rapidity at RHIC and 18\% at the LHC.

\begin{figure}[th!]
    \centerline{
    \includegraphics[width=8.6 cm]{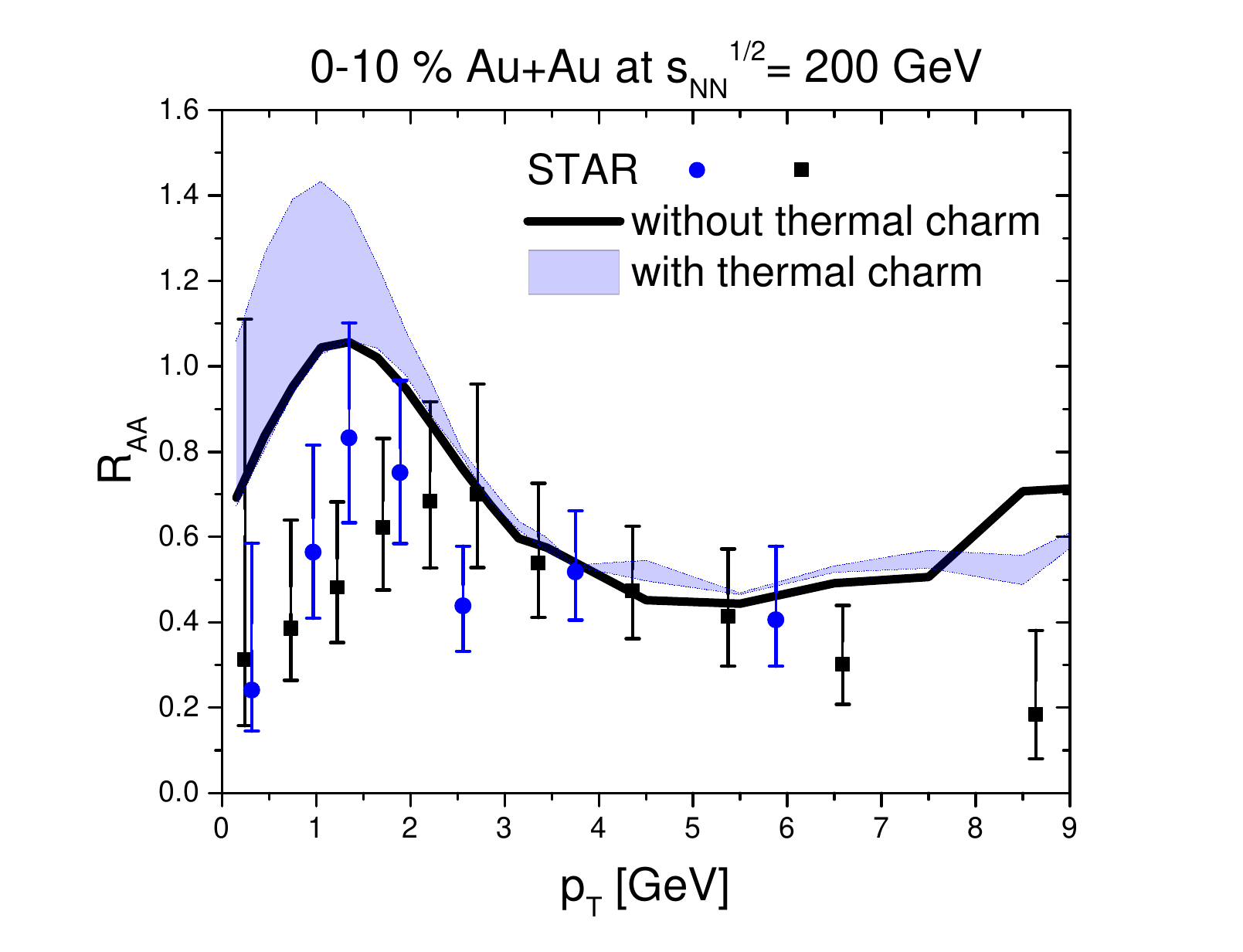}}
    \centerline{
    \includegraphics[width=8.6 cm]{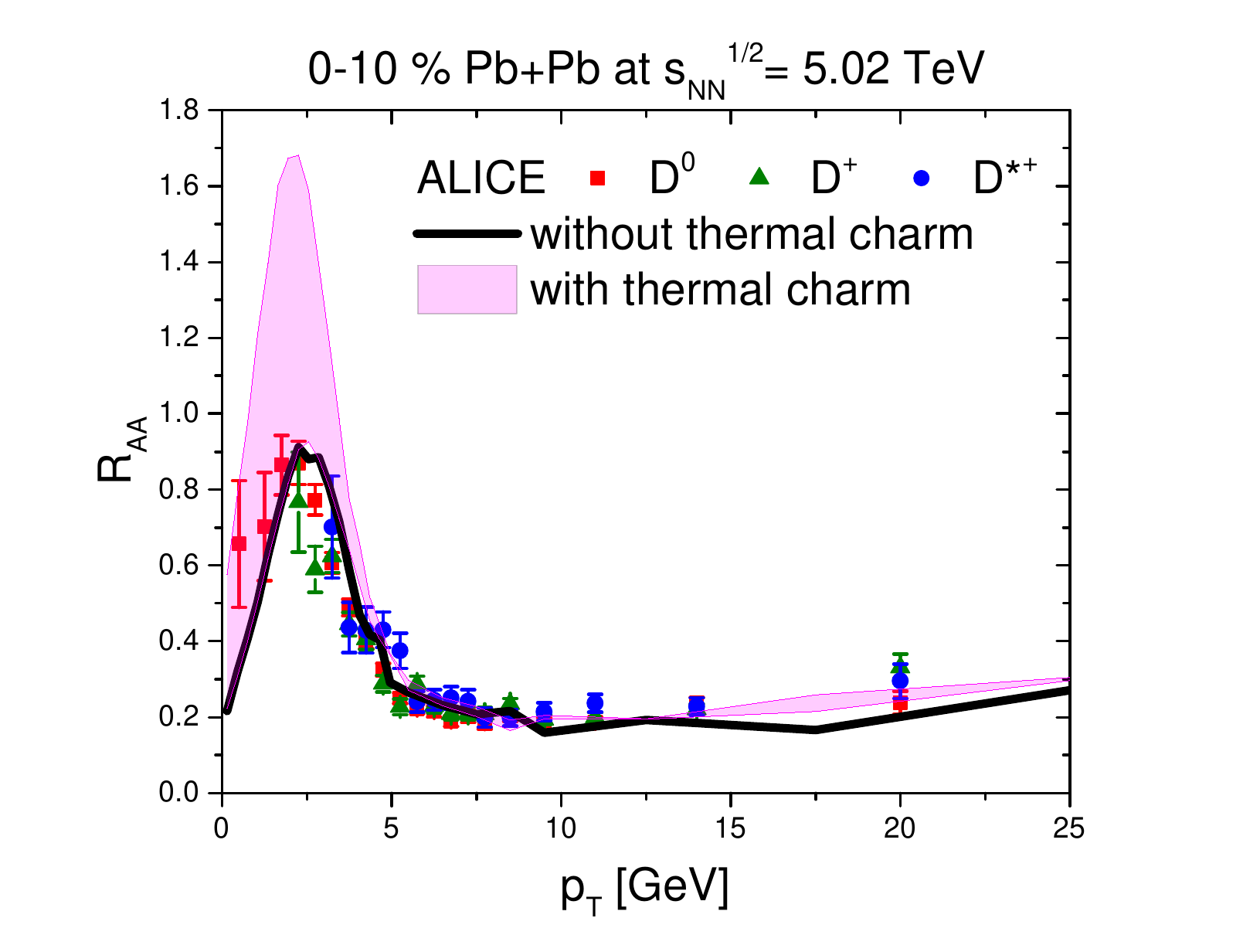}}
    \caption{
        The PHSD results for the $R_{\rm AA}$ of $D$ mesons at midrapidity as a function of transverse momentum (upper) in central Au+Au collisions at $\sqrt{s_{\rm NN}}=$ 200 GeV and (lower) in central Pb+Pb collisions at $\sqrt{s_{\rm NN}}=$ 5.02 TeV with and without thermal charm production in comparison to the experimental data from the STAR~\cite{STAR:2014wif,STAR:2018zdy} and ALICE~\cite{ALICE:2021rxa} Collaborations. The charm quark mass for thermal production varies from 1.2 GeV to 1.8 GeV, which corresponds, respectively, to the upper and lower limits of the colored bands.
    } 
    \label{raa-fig}
\end{figure}

Finally, we examine the effects of thermal charm -- including the interactions of charm quarks in the QGP -- on experimental observables in heavy-ion collisions such as the nuclear modification of $D$ mesons $\rm R_{AA}$, which is defined as
\begin{equation*}
    R_\mathrm{AA}(p_T)\equiv\frac{dN_D^{\rm A+A}/dp_T}{N_{\rm binary}^{\rm A+A}\times dN_D^{\rm p+p}/dp_T},
\end{equation*}
where $N_D^{\rm A+A}$ and $N_D^{\rm p+p}$ are, respectively, the number of $D$ mesons produced in heavy-ion collisions and in p+p collisions, and $N_{\rm binary}^{\rm A+A}$ is the number of binary nucleon-nucleon collisions in heavy-ion collision for the considered centrality class.

Figure \ref{raa-fig} shows the $R_{\rm AA}$ of $D$ mesons including thermal charm production at midrapidity as a function of transverse momentum (upper) in central Au+Au collisions at $\sqrt{s_{\rm NN}}=$ 200 GeV and (lower) in central Pb+Pb collisions at $\sqrt{s_{\rm NN}}=$ 5.02 TeV, compared to the experimental data from the STAR~\cite{STAR:2014wif, STAR:2018zdy} and ALICE~\cite{ALICE:2021rxa} collaborations. So far, the charm quark mass has been taken to be 1.5 GeV. However, the thermal production of charm pairs depends on the charm quark mass due to the threshold of $2m_c$. If the charm quark is lighter, its thermal production will increase. We vary the charm quark mass from 1.2 GeV, which is close to the bare mass in QCD, 1.8 GeV, which is close to the $D$ meson mass. The upper limit of the bands in Fig. \ref{raa-fig} corresponds to the $R_{\rm AA}$ of the $D$ meson with the thermal charm quark mass being 1.2 GeV, and the lower limits to 1.8 GeV. As shown in Fig.~\ref{raa-fig}, the thermal charm quarks enhance $R_{\rm AA}$ at small transverse momentum and have only a small effect at large transverse momentum. We note that the discrepancies at large $p_T$ between the solid line and colored band are not the effects of thermal charm quarks but merely statistical fluctuations. The number of thermal charm quark pairs at midrapidity is 0.7$-$0.053 per unit rapidity at RHIC and $11.5-0.55$ at the LHC for $m_c = 1.2 - 1.8$ GeV.

Based on Fig.~\ref{raa-fig}, it is hard to say something robust about the thermal charm quark effects on $R_{\rm AA}$ at RHIC, as the contribution from the thermal charm is small compared to the uncertainties in the experimental data. On the other hand, the effects of thermal charm quarks are more visible at the LHC because the error bars of the experimental data are relatively small and more thermal charm pairs are produced. The $R_{\rm AA}$ of $D$ mesons including thermal charm for $m_c= 1.8$ GeV is almost the same as the results without thermal production. For $m_c= 1.2$ GeV, however, the effects are clearly visible, and the $R_{\rm AA}$ of $D$ mesons is enhanced at low transverse momentum. When compared with the experimental data from the ALICE Collaboration, the small charm quark mass of 1.2 GeV overestimates the experimental data. This suggests that a larger charm quark mass is more probable for thermal production in heavy-ion collisions, which is consistent with the gaining mass in the QGP through particle dressing~\cite{Moreau:2019vhw, Gubler:2020hft}.

\begin{figure} [t!]
    \centerline{
    \includegraphics[width=8.6 cm]{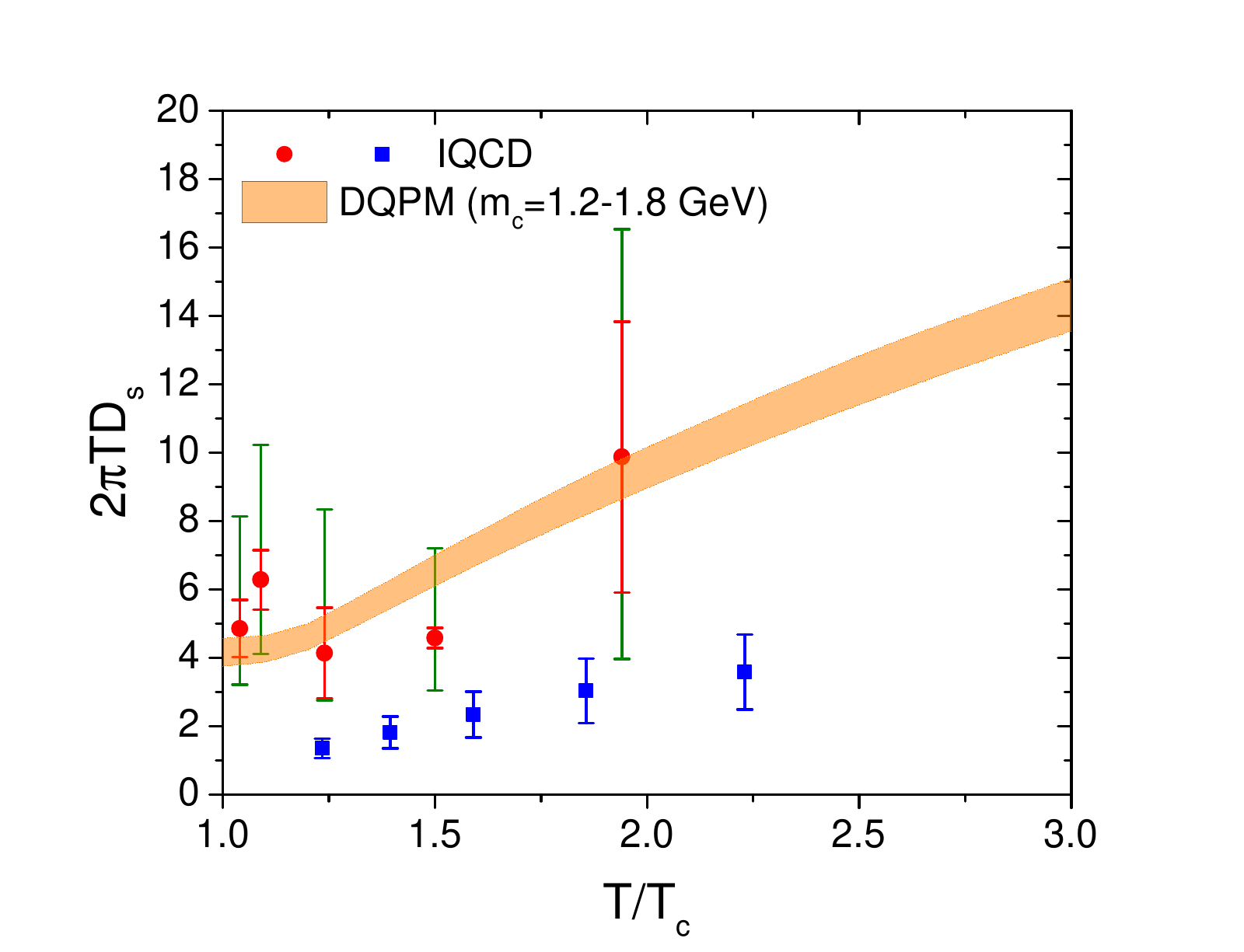}}
    \caption{
        The DQPM results for the scaled spatial diffusion coefficient of charm quarks as a function of the scaled temperature ($T_c=$ 0.158 GeV) for charm quark masses ranging from 1.2 GeV to 1.8 GeV, which correspond to the upper and lower limits of the orange band. The coefficients are compared with those from lattice calculations~\cite{Banerjee:2011ra, Altenkort:2023eav}.
    }
    \label{fig-Ds}
\end{figure}

Finally, we discuss the effect of the charm quark mass on its spatial diffusion coefficient in the QGP. Figure \ref{fig-Ds} shows the spatial diffusion coefficient of charm quarks as a function of temperature $T$ scaled by the critical temperature $T_c$. The upper and lower limits of the orange band correspond to $m_c=$ 1.2 GeV and 1.8 GeV, respectively. The diffusion coefficient decreases with increasing charm quark mass. The orange band is compared with two lattice calculations~\cite{Banerjee:2011ra,Altenkort:2023eav}, one of which is consistent with our results, while the other is about a factor of two lower. A smaller diffusion coefficient can be obtained by increasing the strong coupling for charm quark elastic scattering, which will enhance the thermal production of charm quarks. Therefore, this strongly supports our conclusion that the experimental data favor a heavier charm quark than its bare mass.


Accounting for next-to-leading order $2\to 3$ processes within the non-perturbative DQPM framefork \cite{Grishmanovskii:2023gog} and a comparison to the pQCD results of Ref. \cite{Zhang:2007yoa}
would be an interesting subject for the next study. 
Now we only can speculate that the  radiative processes will enhance charm production in heavy-ion collisions and consolidate our finding that the experimental data favor more  massive charm quarks in the QGP.

\section{Summary}
\label{sec:summary}

We have studied the thermal production of charm quark pairs in the sQGP in central heavy-ion collisions at RHIC and LHC energies, considering two production channels -- quark-antiquark annihilation and two-gluon fusion (as well as their backward reactions). The cross sections are obtained using the DQPM, which reasonably describes the spatial diffusion coefficient of heavy quarks determined by lQCD. Since the elastic scattering for the spatial diffusion coefficient is closely related to charm quark pair production, our results are expected to be reliable.

The thermal charm production has been investigated in heavy-ion collisions using the PHSD, which is a non-equilibrium microscopic transport approach for describing the dynamics of strongly-interacting hadronic and partonic matter produced in heavy-ion collisions. It has two advantages over hydrodynamical approaches, which are widely used for the description of heavy-ion collisions. The first one is that the PHSD does not need to introduce an initial thermalization time, which is necessary for hydrodynamical simulations, but leads to a large uncertainty in the study of thermal charm production. Secondly, the PHSD includes non-equilibrium effects, unlike hydrodynamics, which assumes local thermal equilibrium. This point is also important for the study of thermal charm quark production, since the initial stage of partonic matter, where most thermal charm quarks are produced, is highly off-equilibrium. In the PHSD, two-gluon fusion is dominant in the initial stage and then quark-antiquark annihilation becomes more important when approaching $T_c$, although the production rates from the two channels are similar in a thermalized QGP.

Comparing the initial charm quarks produced through nucleon-nucleon hard scattering and thermal charm quarks, the latter spectrum is softer, and its numbers amount to approximately 6.5\% and 18\% of the initial charm quarks at midrapidity in central heavy-ion collisions at RHIC and LHC energies, respectively, for $m_c$= 1.5 $\rm{GeV}$. However, these numbers increase to 30\% and 92\% at RHIC and the LHC, respectively, for $m_c$= 1.2 $\rm{GeV}$, and decrease to 2.3\% and 4.5\%, respectively, for $m_c$= 1.8 $\rm{GeV}$. In this way, the thermal production of charm quarks strongly depends on the charm quark mass. If the charm quark mass is as small as the QCD bare mass, more charm quarks are produced, and, assuming the charm quark mass to be close to the $D$ meson mass, the thermal production is strongly suppressed.

Compared with the experimental data on the $R_{\rm AA}$ of $D$ mesons in heavy-ion collisions, especially at LHC, heavier charm quark masses are favored. This is consistent with the idea that the heavy quark is dressed and becomes more massive in the QGP.

\section*{Acknowledgements}

The authors acknowledge inspiring discussions with J. Aichelin, W. Cassing, and C. Greiner. Furthermore, we acknowledge support by the Deutsche Forschungsgemeinschaft (DFG, German Research Foundation) through the grant CRC-TR 211 'Strong-interaction matter under extreme conditions' - Project number 315477589 - TRR 211. This work is also supported by the European Union’s Horizon 2020 research and innovation program under grant agreement No 824093 (STRONG-2020). The computational resources have been provided by the LOEWE-Center for Scientific Computing and the "Green Cube" at GSI, Darmstadt, and by the Center for Scientific Computing (CSC) of the Goethe University, Frankfurt.

\bibliography{main}

\end{document}